\documentclass[a4paper,8pt]{article}
\usepackage[cp1251]{inputenc}
\usepackage[russian,english]{babel}
\usepackage{array,longtable}
\usepackage{amssymb, latexsym, amsthm, amsmath, amsxtra}
\usepackage{graphicx}

\graphicspath{{}}
\DeclareGraphicsExtensions{.pdf,.png,.jpg,}

\makeatletter

\begin{document}
\large

\begin{center}
\title{}{\bf On the Soliton Model of a Point Particle Spin  }
\vskip 1cm

\author{}
 R.K. Salimov \textsuperscript{1}, E.G. Ekomasov \textsuperscript{1,2,3}
{}

\vskip 1cm

{ \textsuperscript{1} Bashkir State University, 450076, Ufa, Russia }

{ \textsuperscript{2} University of Tyumen, Tyumen, 625003 Russia}

{ \textsuperscript{3} South Ural State University (National Research University), Chelyabinsk, 454080 Russia}

\vskip 0.5cm
e-mail: salimovrk@bashedu.ru

\end{center}

\vskip 1cm

{\bf Abstract}
 \par
 A system consisting of a point material particle and a scalar field described by the nonlinear Klein-Gordon equation has been considered. It has been shown that, when taking into account relativistic effects, in the case of small rest masses of a particle an energy minimum at zero velocity is impossible for such a particle.  It has been proved that under certain parameters the spin state of a particle has a lower energy than its steady state does. Such a behavior is interesting for the construction of soliton models of spin. By a point particle spin an intrinsic angular momentum (orbital, in strict sense, but with a short orbital radius) of a point particle is meant.

 \par
 \vskip 0.5cm


\par
\vskip 1cm

In modern nonlinear wave theory, great attention is focused on the determination of spatially localized and time-periodic solutions for numerous models and various dimensions [1-6]. Such spatially localized solutions have a finite energy and correspond to static particle-like objects or various travelling waves. One of the invariant nonlinear differential equations studied most often is the Klein-Gordon equation, the sine-Gordon, in particular. It has a lot of applications in different fields of physics including hydrodynamics, condensed matter physics, field theory, etc. The Klein-Gordon equations are Lorentz-invariant and their solutions have  relativistic effects [7]. Paper [8] studies the nonstationary inhomogeneity of the spatially one-dimensional Klein-Gordon equation, which was considered as the soliton model of the system with the undamped motion of the particle. The two-dimensional case should be studied to consider such a system as a model with  the undamped spin. The paper illustrates that under certain parameters in the 2D and 3D cases the energy minimum is impossible for a stationary localized particle.
	To consider nonstationary inhomogeneities let an inhomogeneity be produced by a point particle with the rest mass m, the coordinates of which are denoted as $x_1,y_1, v_x=\dot x_1, v_y=\dot y1$. The total energy or the Hamiltonian of such a system can be written in the form:
\begin{align}
 H=H_{def}+H_u+H_{int} \label{eq:1}
   \end{align}

where H def is the energy of the particle creating the inhomogeneity
 \begin{align}
 H_{def}=\frac{m}{\sqrt{(1-v^2)}}  \label{eq:1}
   \end{align}
H u is the energy of the scalar field
  \begin{align}
  H_u= \int\limits_{-\infty}^{\infty}\int\limits_{-\infty}^{\infty} (\frac{u_x^2}{2}+\frac{u_y^2}{2}+\frac{u_t^2}{2}+V(u)) dxdy  \label{eq:3}
   \end{align}

H int is the energy of interaction between the scalar field and the particle

\begin{align}
  H_{int}=\int\limits_{-\infty}^{\infty}\int\limits_{-\infty}^{\infty}  q(x_1,y_1,v_x,v_y,x,y)W(u) dxdy  \label{eq:3}
   \end{align}

The function V(u) in equation (3) was written as follows:
 \begin{align}
  V(u)=\frac{u^2}{2}+\frac{u^4}{2}  \label{eq:3}
   \end{align}

The scope of potential $q(x_1,y_1,x,y,v)$  is bounded by an ellipse. Here, for simplicity, we assume that the particle moves in the direction of  X-axis.

\begin{align}
  q(x,y,x_1,y_1,v_x,v_y))=\frac{U_0}{\sqrt{1-v^2}}
  \nonumber\\if (R^2-\frac{(x-x_1)^2}{1-v^2}-(y-y_1)^2)\geq 0
  \end{align}
\begin{align}
  q(x,y,x_1,y_1,v_x,v_y))=0
  \nonumber\\if (R^2-\frac{(x-x_1)^2}{1-v^2}-(y-y_1)^2)< 0
  \end{align}

In the case of a stationary localized particle the energy minimum $H_u+H_def$  is achieved at the stationary solution of the equation

\begin{align}
  u_{xx}+u_{yy}-u_{tt}=u+2u^3-\nonumber\\U_0 \sin(u) q(x_1,y_1,v_x,v_y,x,y)
   \end{align}

The equation has a numerical stationary cylindrically symmetric solution for, for example, the parameters $U_0=20, R=2 $. For further reasoning it is enough to know that the solution is monotone, i.e. decreases with growth $r=\sqrt{(x-x_1)^2+(y-y_1)^2}$  and $|u|_{max}<\pi$. Here, the coordinates of the center of the potential $x_1,y_1$ coincide with the coordinates of the center of the soliton solution $x_0,y_0$.

We then assume that at small rest masses $m$  for a stationary localized particle the energy minimum  $H_{def}+H_u+H_{int}$ is achieved. If the energy minimum is achieved when the solution is in the form of a stationary soliton, then the minimum cannot be further decreased. Let us show it is not true. For that we assume that the particle rotates around the center of the soliton solution at some  $v$ velocity while being at the distance $L$ from the center of the soliton solution $x_0,y_0$. With regard for retardation the scope of the potential $q$ will have the form shown schematically in Fig. 1

\begin{figure}[h]
\center
\includegraphics[width=6cm, height=6cm]{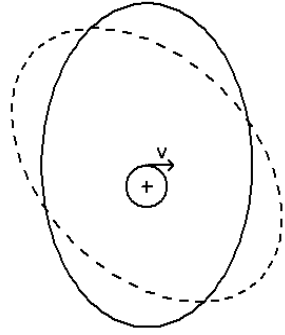}
\caption{ The schematic illustration of $q$ potential which is oblate in the direction of movement. The ellipse bounding the instantaneous potential is showed by the full lines; the dash lines show the area of  the action of the potential with regard for retardation. The cross marks the center of the soliton solution. The smaller circle around the cross marks the trajectory of the $q$ potential center movement with respect to the soliton solution center.  }
\label{schema}
\end{figure}

 We now send $L$ to 0 without a decrease in velocity $v$. In the expression  $H_{def}+H_u+H_{int}$  the energy $H_u$ stays the same while the energy $H_{def}+H_{int}$ can decrease at small $m$ and $v>0$. Indeed, when considering the expression
\begin{align}
 \frac{m}{(1-v^2)^{(1/2)}}
  +\nonumber\\ \frac{U_0}{(1-v^2)^{1/2}} \int\limits_{y_1-R}^{y_1+R}dz \int\limits_{x_1-w}^{x_1+w}(2cos^2(u/2))dx
    \end{align}
where
\begin{align}
w(y,y_1,v)=\sqrt{(1-v^2)(R^2-(y-y_1)^2)}
 \end{align}
$x_0=x_1$;$y_1=y_0$

it can be seen that that the second addend gets smaller at $v>0$ in (25) since the region of integration over $x$ with the length $2w$ decreases proportionally to the factor $\sqrt{1-v^2}$, and even though each addend in the  integral sum increases proportionally to the factor $1/\sqrt{1-v^2}$ , the integral decreases due to the monotone increase of the value $2cos^2(u/2)$ with increasing $(x-x_1)^2$. When denoting the decrease of the  augend in (25) as $\delta$ we obtain the energy  $H_{def}+H_{int}$  which can be decreased at  $m/\sqrt{1-v^2}<\delta$. Then the minimum energy at the stationary position is not achieved for small $m$. The conclusion can also be generalized for the 3D case.
In other words, at a small rest mass $m$ at the velocity of $v=0$ and $U_0$, $R$ for the system presented there is no stable minimum energy state such that has a stationary solution in the form of a soliton. Since the conclusion is valid for all inertial reference frames, we obtain that in the state with the minimum energy the steady motion of the particle at a constant velocity is also impossible. That is, a stable state for a particle in such a system is, in the general case, some accelerated movement.
   Thus the model presented describes a particle with a small rest mass permanently moving at a certain acceleration. In the case of the finite motion of the particle in the two- and three-dimensional cases, the system will have  some nonzero angular momentum. In such a case it can be considered as a soliton model of spin.


\end{document}